\documentclass[preprint,preprintnumbers,amsmath,amssymb]{revtex4}

\usepackage{graphicx}
\usepackage{subfigure}
\usepackage{dcolumn}
\usepackage{bm}

\begin{document}

\title{\bf Adsorbate-adsorbate interactions and chemisorption at different coverage studied by accurate {\em ab initio} \bf calculations: CO on transition metal surfaces}

\author{Sara E. Mason, Ilya Grinberg  and Andrew M. Rappe}
\affiliation{
The Makineni Theoretical Laboratories, Department of Chemistry\\ University of Pennsylvania, Philadelphia, PA 19104-6323
}%

\date{\today}

\begin{abstract}We use density functional theory (DFT) with the generalized gradient approximation (GGA) and our first-principles extrapolation 
method for accurate chemisorption energies {[Mason {\em et al.},
Phys.\ Rev.\ B {\bf 69}, 161401R (2004)]} to calculate the
chemisorption energy for CO on a variety of transition metal
surfaces for various adsorbate densities and patterns.  We
identify adsorbate through-space repulsion, bonding competition, and
substrate-mediated electron delocalization as key factors determining preferred
chemisorption patterns for different metal surfaces and adsorbate 
coverages.  We discuss how the balance of these interactions, along
with the inherent adsorption site preference on each metal surface, can
explain observed CO adsorbate patterns at different
coverages.

\end{abstract}

\maketitle

\section{\label{sec:intro}Introduction }

The prototypical CO/transition metal adsorption systems, though
extensively studied, continue to be of significant contemporary
research
interest.~\cite{Besley05p184706,Mitsui05p036101,Yamagishi05p8899}
These systems owe their continued examination in part to the
significant variations in
CO bonding properties from one transition metal
to another, such as the evolution of chemisorption geometry with
adsorbate coverage $\Theta$.  Attempts to describe adsorption
trends as a function of $\Theta$ are complicated by the many phenomena
involved, such as through-space interactions between adsorbates,
substrate-mediated interactions, surface relaxations and
reconstructions induced by adsorption, adsorption at different
high-symmetry sites, patterned overlayer formation, and transitions
between different adsorbate arrangements on the surface under
different conditions.~\cite{Hammer01p205423,Linke01p8209,Persson90p5034,Gurney87p6710}

There are many studies that address
adsorbate-adsorbate
interactions~\cite{Brako00p185,vanBavel03p524,Kato02p239,Zaera02p4043,Kose99p8722,Osterlund99p4812,Kevan98p19,Wei97p49,Persson90p5034},
and it is beyond the present scope to review these and other studies
in full.  One major conclusion is that properties of sub-monolayer CO
overlayers such as chemisorption energy and surface site occupation
are found to be significantly coverage dependent.  Interactions
between CO molecules are strongly repulsive at short range; beyond
nearest neighbor distance, inter-adsorbate interactions are
non-monotonic and vary as a function of adsorbate coverage.  A better
understanding of adsorbate interactions would be helpful for
understanding the wide variety of adsorption systems of interest to
the surface science
community.~\cite{Ge04p1551,Liu03p1958,Zaera02p4043,Mar94p188,Tarek99p964,Yourdshahyan01p081405}

To understand how chemisorption evolves with $\Theta$, it is necessary
to identify the factors influencing the adsorbate-adsorbate
interactions, and how these factors vary from metal to metal.  For
this purpose, it is advantageous to have a data set spanning a variety
of metals, and with a broad range (in terms of coverage, site
occupancy, and overlayer pattern) of chemisorption geometries.  With
such a dataset, chemisorption energies between different CO overlayer
structures can be compared for different metals, elucidating trends
and revealing important interactions.

The goal of comparing chemisorption at different coverages, and at
different adsorbate overlayer patterns for a given coverage, is well
suited for theoretical study.  We can
model adsorption at coverages and in patterns not experimentally
observed.  However, it is necessary to have a means for accurately
modeling chemisorption on a variety of metals, and for all
adsorption sites, coverages, and overlayers of interest.  Owing to
advances in computer power and algorithms, many aspects of surface
science can be quantitatively explained through theoretical studies at
the DFT-GGA level.~\cite{Gross03, Greeley02p319}  

Hindering the use of
DFT for quantitative investigation of CO/metal adsorption systems is
the tendency of DFT calculations to overestimate chemisorption
energies and to favor multiply coordinated adsorption sites over the
top site, in disagreement with experimental data.  This ``CO/metal
puzzle'' was first comprehensively addressed by Feibelman {\em et
al.}\cite{Feibelman01p4018}  Subsequently,
several studies have traced the CO/metal
puzzle to its
origin.~\cite{Grinberg02p2264,Gil03p71,Kresse03p073401}  

Following
these investigations, we recently derived a first-principles
extrapolation procedure~\cite{Mason04p161401R} which, when applied to
eight different single-crystal substrates, reduced over-binding and
achieved site preferences consistent with experimentally observed
adsorption in all cases.  Other works discussing how to correct for
the DFT site-preference problem include Olsen {\em et
al.}~\cite{Olsen03p4522} and Kresse {\em
et al.}~\cite{Kresse03p073401}  In the work of Olsen {\em et al.},
relativistic effects are found to be important to correct site
preferences for CO on Pt~(111).  However, this fails to explain why DFT
finds incorrect site preference on non-relativistic Cu, as recently
also pointed out by Gajdo\v{s} {\em et al.}~\cite{Gajdos05p117}
Kresse and co-workers employ a GGA$+U$ type exchange-correlation energy
functional. However, the correct value of $U$ is not known {\em a
priori}.  In contrast, our method uses {\em ab initio} calculations to
unambiguously calculate the correction magnitude.  The present work is
the first time any of these three accurate methods has been applied
to multiple metals, surfaces, and coverages in a single study.

\section{\label{sec:method} Methodology and Computational Details}    

In the present study, we investigate chemisorption of CO on transition
metals as a function of metal, surface termination, adsorption site,
and CO overlayer pattern.  The focus of the present
work is transition metal chemisorption, but some
additional results for CO on Al~(111) and (100) surfaces are also
presented  for the purpose of contrasting transition metals with
an $sp$ metal.

DFT calculations are carried out using
the PBE GGA exchange-correlation functional~\cite{Perdew96p3865} and
norm-conserving optimized pseudopotentials~\cite{Rappe90p1227} with
the designed nonlocal method for
metals.~\cite{Ramer99p12471,Grinberg01p201102}  All pseudopotentials
were designed using the OPIUM pseudopotential package.~\cite{Opium}
Metal surfaces are modeled as $c(4\times2)$ or $p(2\times2)$ slabs of
five layers separated by vacuum, with atomic relaxation allowed in the
two topmost layers.  
Calculations are done, and $E_{\rm chem}$ tested to be converged within
less than 0.025~eV, using a $4\times4\times1$ grid of irreducible
Monkhorst-Pack $k$-points.~\cite{Monkhorst76p5188}

All values for $E_{\rm chem}$ reported here have been corrected using
our first-principles extrapolation procedure.~\cite{Mason04p161401R}
We have tested that the corrections to $E_{\rm chem}$ do not
appreciably evolve with $\Theta$, affecting corrected values for
$E_{\rm chem}$ by less than 0.01~eV/molecule.  
CO chemisorption is modeled at coverages of
$\Theta$=0.25, $\Theta$=0.5, $\Theta$=0.75, and $\Theta$=1~ML (monolayer).

In these calculations, the CO bond is
held perpendicular to the surface.  Eliminating the CO tilt
angle as a degree of freedom facilitates comparison of chemisorbed
and gas-phase adsorbate-adsorbate
interactions, as well as comparison of CO-CO
interactions over different metals.  None of the higher-coverage
chemisorption structures on the (100) surfaces were found to have
lateral forces on CO under this constraint.  In some of the 
higher-coverage structures on the (111) surfaces, lateral forces 
of $\approx$0.01--0.05~eV/\AA\ were found on the C and O atoms, as
indicated in Table~\ref{table:111Results}.  No dipole correction
scheme to decouple the periodic slabs was included, as such
corrections were found to affect the total energy of the chemisorption
systems by less than 0.01~eV/cell.

Values for $E_{\rm chem}$ are calculated as
\begin{eqnarray}
E_{\rm chem} = (-E_{\rm surface-CO} + NE_{\rm CO} + E_{\rm surface})/N
\end{eqnarray}

\noindent where $N$ is the number of CO molecules per unit cell.
With this definition, chemisorption energies are reported in eV/CO
molecule, and positive values of $E_{\rm chem}$ represent favorable
adsorption on the surface.  

Note: The overlayer naming convention denotes both coverage and sites:
one letter for $\Theta$=0.25, two for $\Theta$=0.5, three
for $\Theta$=0.75, and four letters for $\Theta$=1.

\begin{figure}
     \centering
     \subfigure[]{
          \label{fig:111Quarter}
          \includegraphics[height=1.15in]{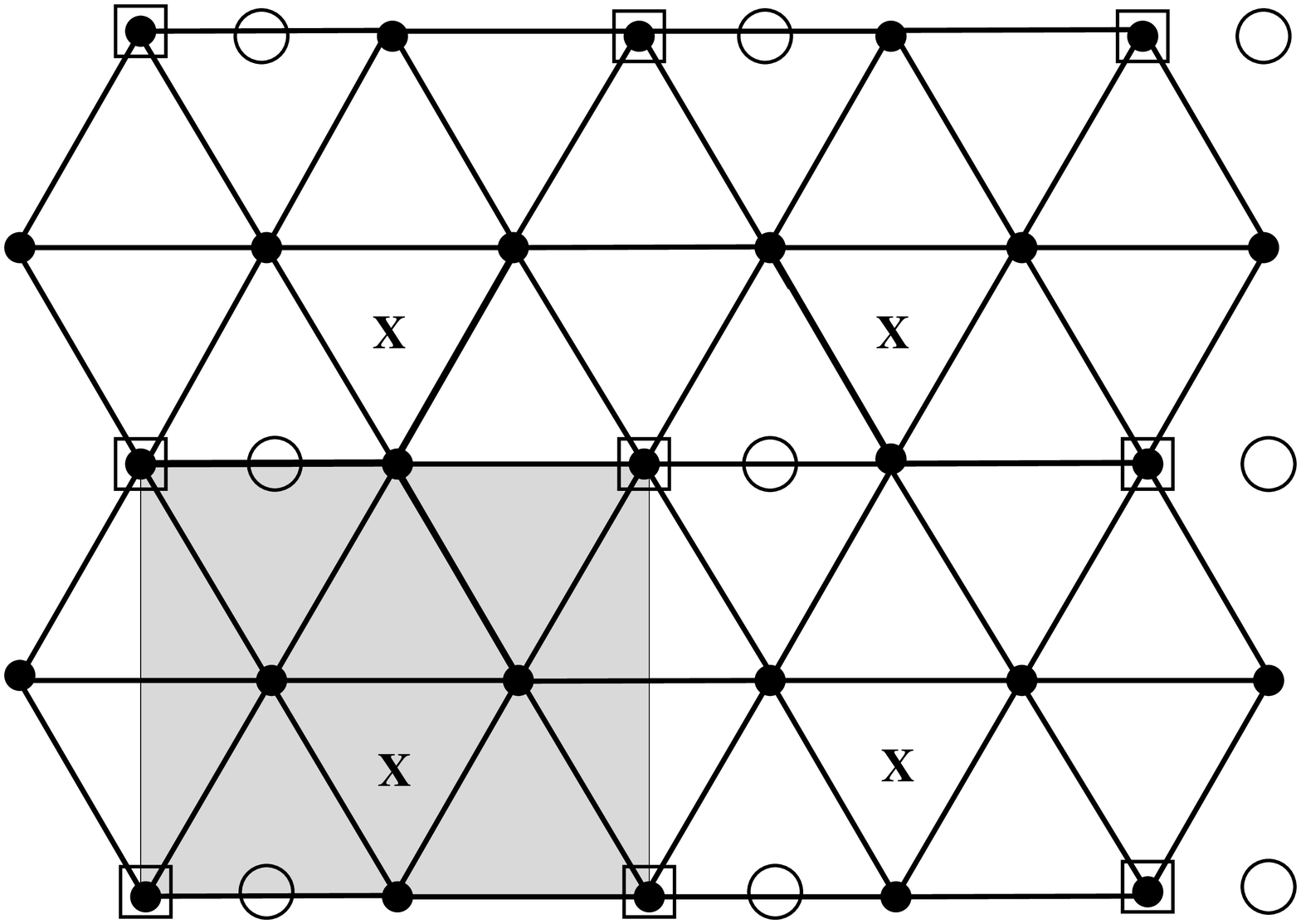}}
     \subfigure[]{
          \label{fig:111TH}
          \includegraphics[height=1.15in]{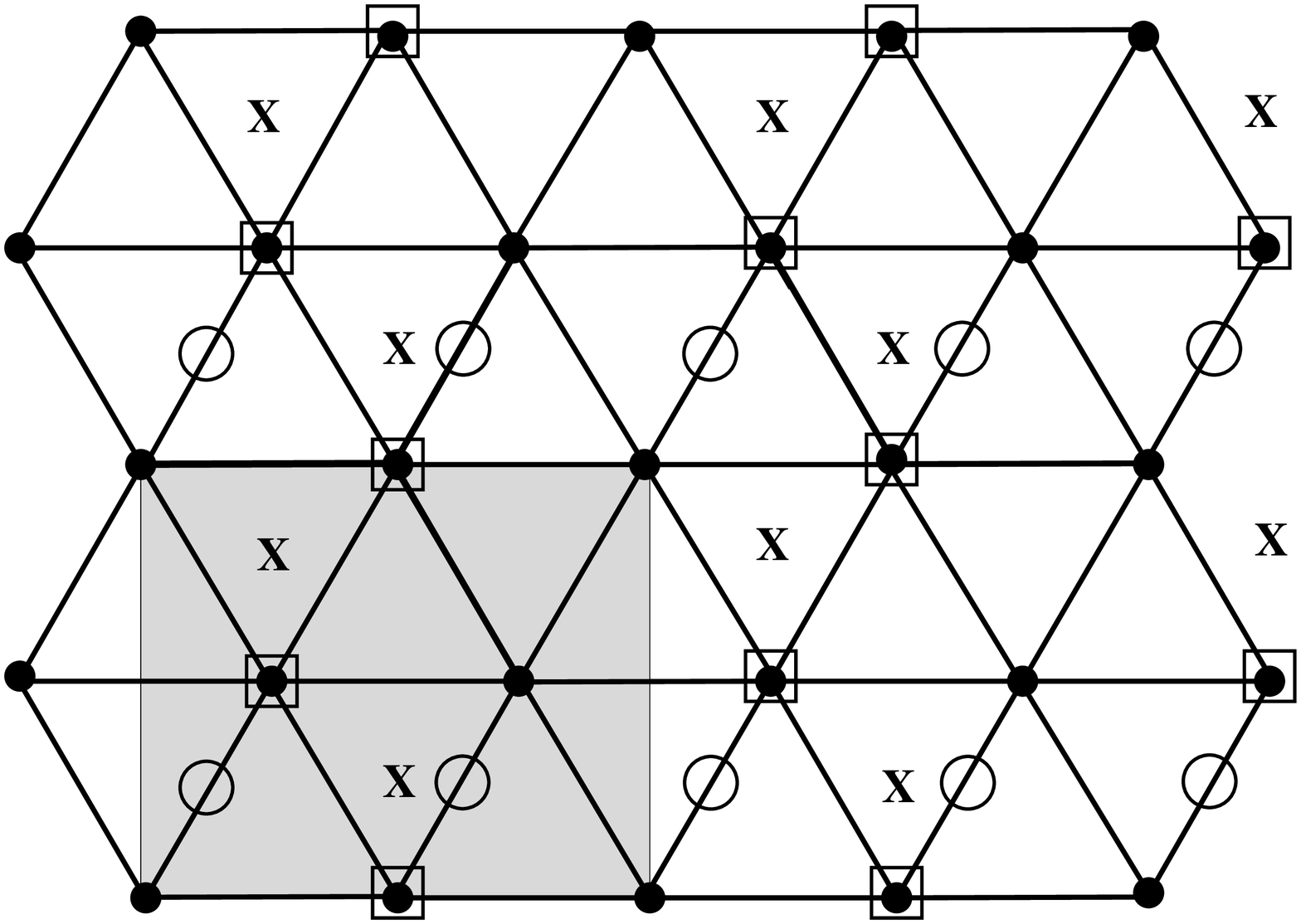}}\\
     \vspace{0.2in}
     \subfigure[]{
           \label{fig:111Bb2}
           \includegraphics[height=1.15in]{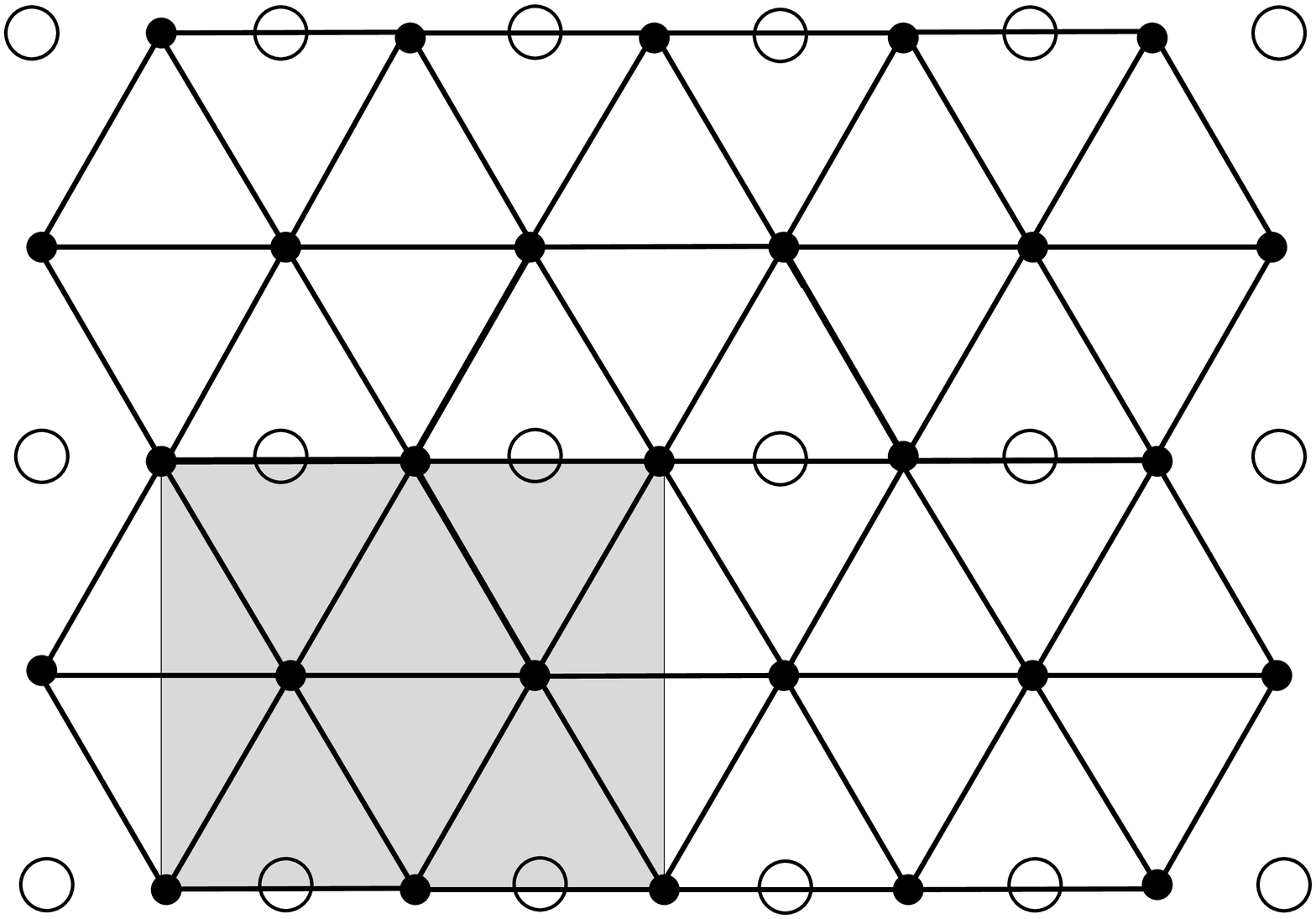}}
     \subfigure[]{
           \label{fig:111Bb3}
          \includegraphics[height=1.12in]{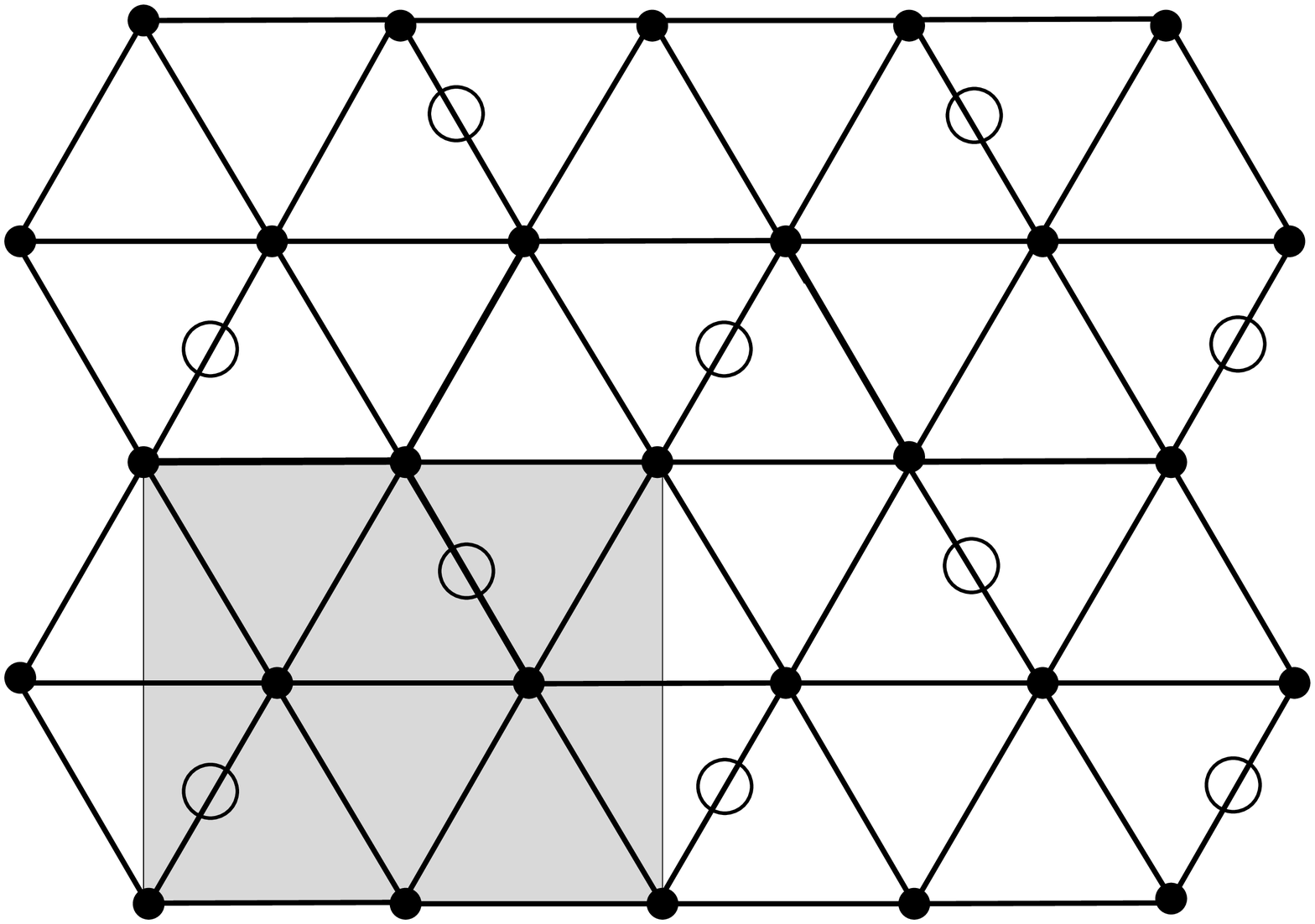}}
\caption{(a) Schematic of (111) surfaces with CO at $\Theta$=0.25.
Overlayer t is indicated by squares, b by circles, and h by ``X''.
The cell is indicated by the shaded region. (b) Schematic of (111) surface with
CO at $\Theta$=0.5 in overlayer tt (squares), in the bb1 overlayer
pattern (circles) and hh pattern (``X''). (c) Schematic of the (111)
bb2 structure.  (d) Schematic of (111) bb3 structure.}
     \label{fig:111Exclusive}
\end{figure}

On the (111) surfaces of Pt, Rh, and Pd, we modeled chemisorption at
$\Theta$=0.25 at top (t), bridge (b), hcp (h) and fcc (f) adsorption sites, as shown
in Figure~\ref{fig:111Quarter}.  We consider our results for
$\Theta$=0.25 to be the low-coverage limit values of $E_{\rm chem}$.
As a test of the validity of this
assumption, we also calculated $E_{\rm chem}$ on each metal at each
site in a $p(2\times3)$ cell, corresponding to a coverage of
$\Theta$=0.17, on both the (111) and (100) surfaces.  The change
in $E_{\rm chem}$ from $\Theta$=0.17 to $\Theta$=0.25
was less than 0.04~eV for all sites and metals, with the same
site preference.  The most important characteristic 
of the t, b, f, and h arrangements
is the absence of nearest
neighbors.  

At $\Theta$=0.5, we modeled chemisorption at
top (tt) and hcp (hh) adsorption sites, as shown in
Figure~\ref{fig:111TH}.  We did not model fcc adsorption at
$\Theta$=0.5 due to the very similar energies for f and h
for all surfaces as shown in Table~\ref{table:111Results}.
The tt and hh structures allow us to
investigate how chemisorption changes when adsorbates are
separated by nearest neighbor distances.  

We model three different bridge site
configurations, bb1, bb2, and bb3, as shown in
Figures~\ref{fig:111TH}, \ref{fig:111Bb2}, and \ref{fig:111Bb3}.
In bb1 and bb2, nearest
neighbor CO molecules are present.  For the bb2 configuration there is
an additional characteristic: each surface metal atom involved in
chemisorption participates in two carbon-metal bonds, leading to bonding
competition.~\cite{Hammer01p205423}  The bb3 configuration does not
have nearest neighbor CO molecules, and is
one of the observed patterns at $\Theta$=0.5 on the
Pd~(111) surface.~\cite{Rose02p48}  Comparing these three bridge
patterns allows us to deduce the energetic costs of
nearest-neighbor adsorbate
interactions and bonding competition for each metal.  

\begin{figure}
     \centering
     \subfigure[]{
          \label{fig:111TB}
          \includegraphics[height=1.15in]{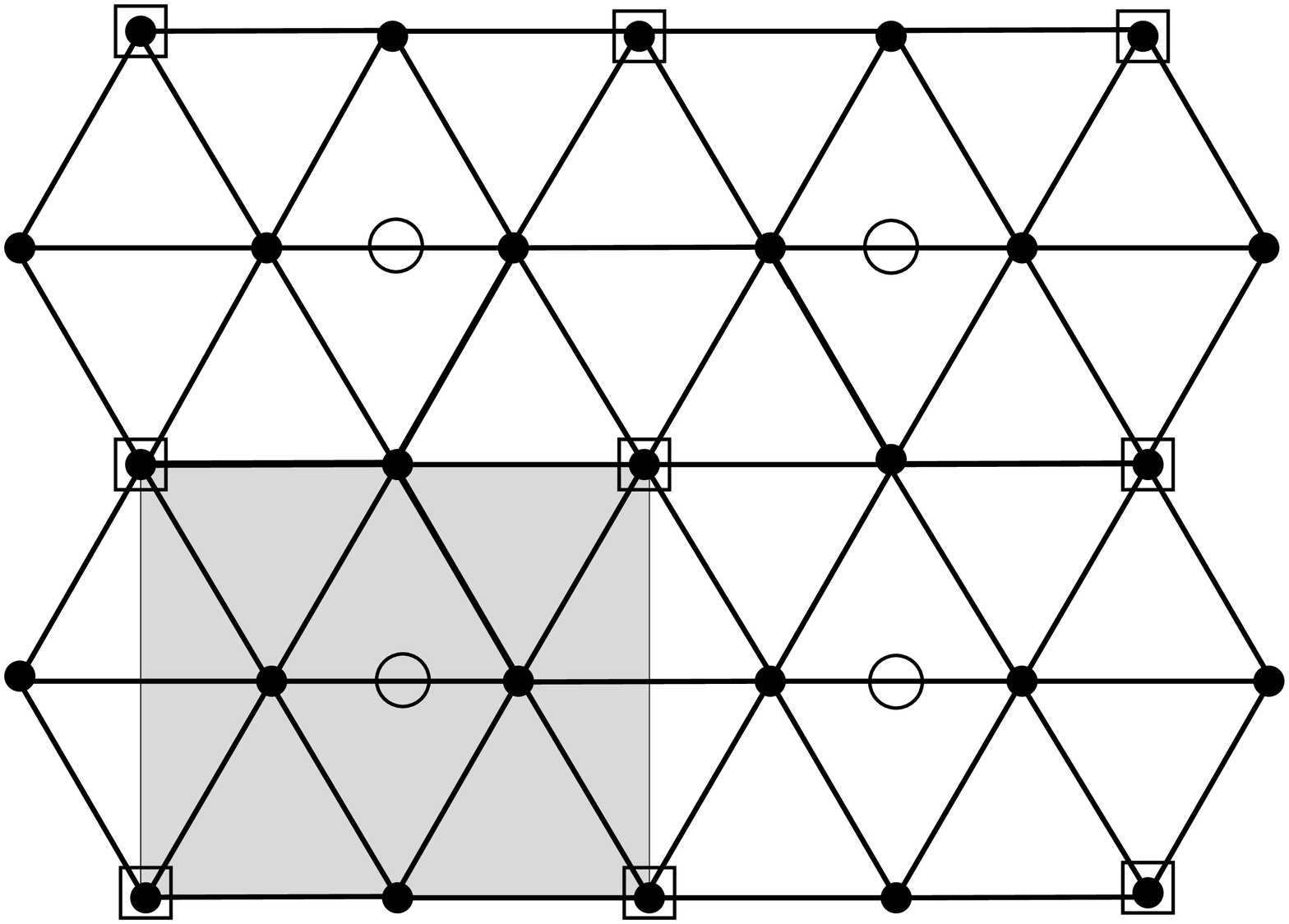}}
     \subfigure[]{
          \label{fig:111p2x2}
          \includegraphics[height=1.15in]{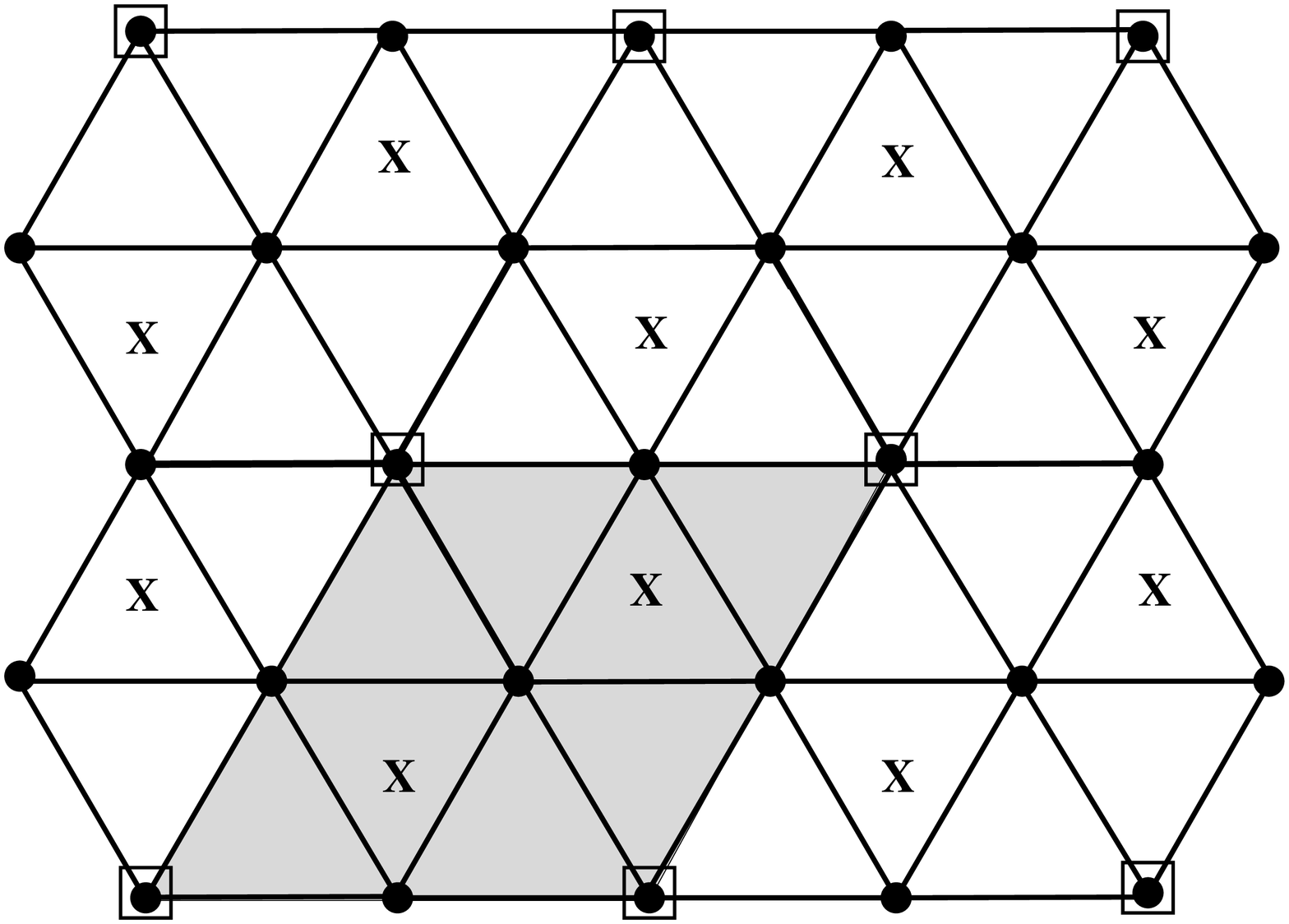}}\\
     \caption{(a) Schematic of (111) tb structure at $\Theta$=0.5.  Occupied top sites are shown with squares, and occupied bridge sites are shown with circles. (b) Schematic of (111) hf and thf structures at $\Theta$=0.5 and $\Theta$=0.75.  In the former, only hcp and fcc sites (``X'') are occupied.  In the latter, the top sites (squares) are also occupied.}
     \label{fig:111Mixed}
\end{figure}

We also considered (see Figure~\ref{fig:111TB}) mixed
occupation of top and bridge (tb) in a $c(4\times2)$ cell,
the observed pattern at
$\Theta$=0.5 on
Pt~(111)~\cite{Pederson99p403,Ogletree86p351,Bondino00pL467}, and mixed
occupation (see Figure~\ref{fig:111p2x2}) of hcp and fcc (hf) in a $p(2\times2)$ cell,
one of the observed patterns on
Pd~(111).~\cite{Rose02p48}  For the hf overlayer, bonding competition
between CO molecules is present.  

At $\Theta$=0.75, we modeled an
overlayer with CO in top, hcp hollow, and 
fcc hollow (thf), Figure~\ref{fig:111p2x2}.  This coverage and pattern
is observed on both Rh and
Pd~(111)~\cite{Beutler98p117,Beutler97p381,Rose02p48}, and is another
configuration where bonding competition is present.  

Chemisorption at
$\Theta$=1 is modeled in a $c(4\times2)$ cell for both top site (tttt) and
hcp (hhhh) adsorption.  Bonding competition cannot exist between top site
adsorbed molecules, but is present in the hhhh configuration.
Consistent with previous work, we find chemisorption at $\Theta$=1 to
be exothermic.  The lack of experimental observation of CO at
$\Theta$=1 is due to kinetic effects.~\cite{Steckel03p085416}

\begin{figure}
     \centering
     \subfigure[]{
          \label{fig:100Q}
          \includegraphics[height=1.2in]{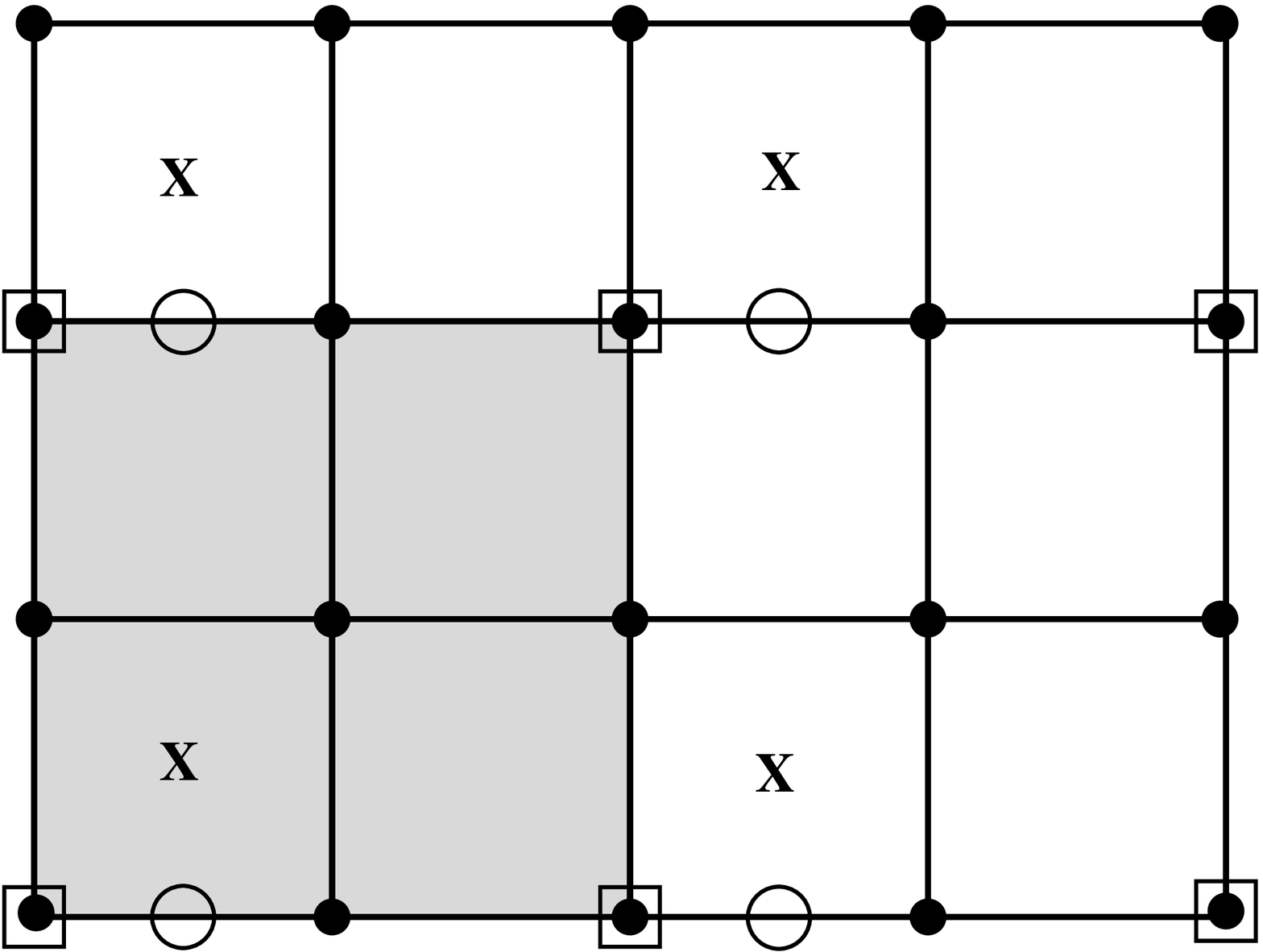}}
     \subfigure[]{
          \label{fig:100Half1}
          \includegraphics[height=1.2in]{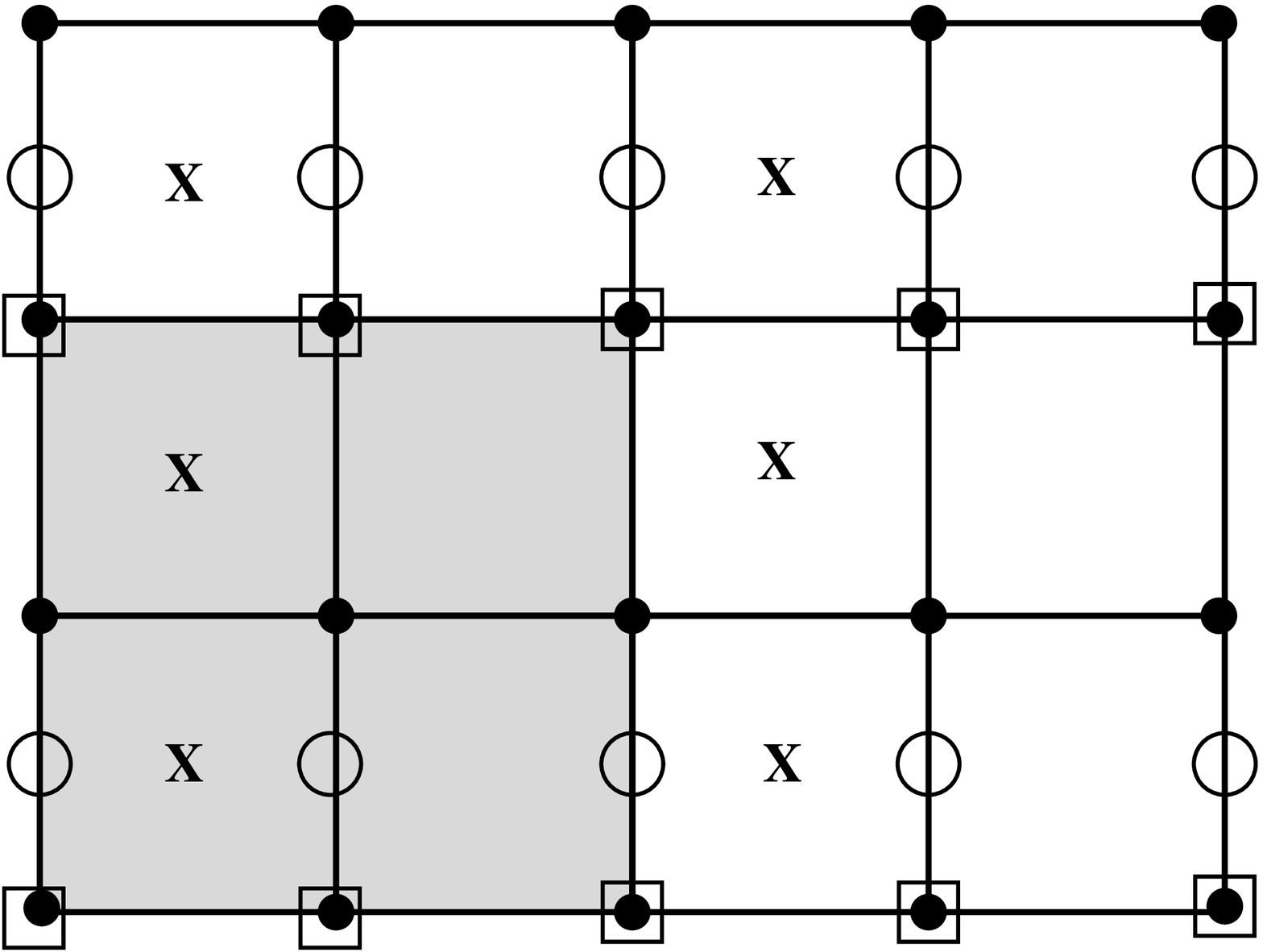}}\\
     \subfigure[]{
           \label{fig:100Half2}
           \includegraphics[height=1.2in]{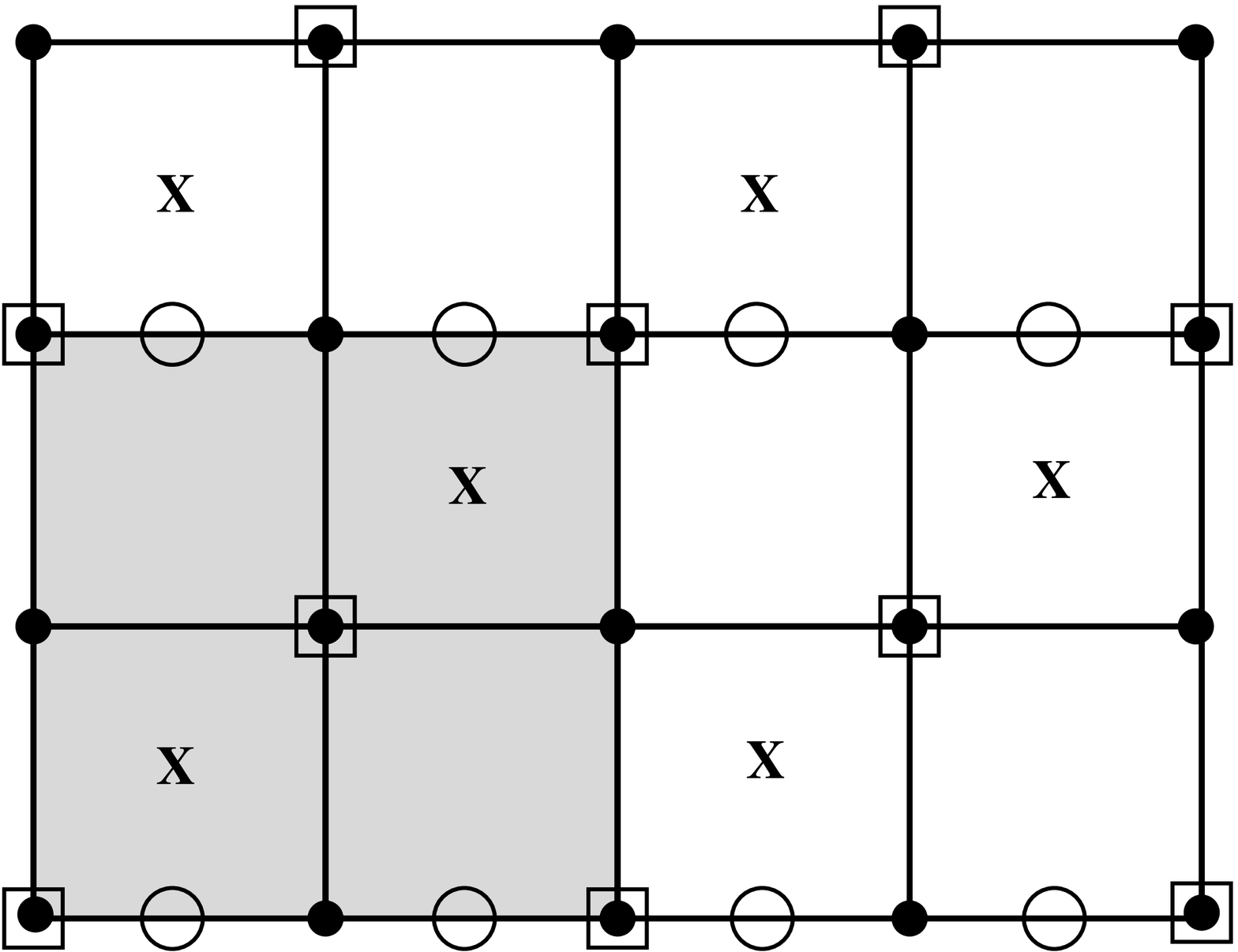}}
     \subfigure[]{
           \label{fig:100Bb3}
          \includegraphics[height=1.2in]{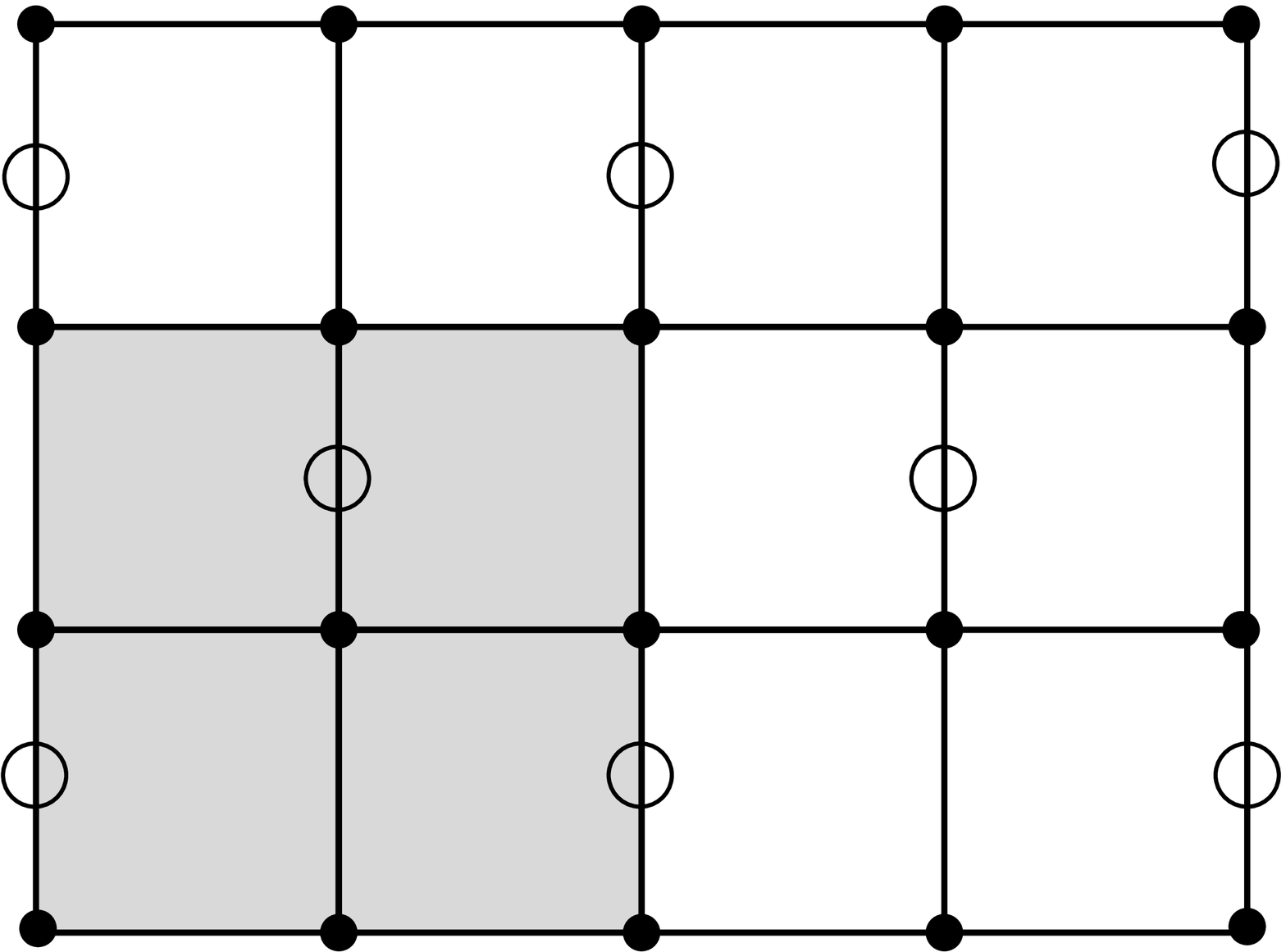}}
     \caption{(a) Schematic of (100) surface with CO at $\Theta$=0.25.  Top site pattern t is indicated by squares, overlayer b by circles, and h by ``X''. (b) Schematic of (100) tt1, bb1, and hh1 overlayers ($\Theta$=0.5).  (c) Schematic of the (100) tt2, bb2, and hh2 overlayers.  (d) Schematic of (100) bb3 structure with CO at $\Theta$=0.5.}
     \label{fig:100Exclusive}
\end{figure}

On the (100) surfaces of Pt, Rh, and Pd, we modeled chemisorption at
$\Theta$=0.25 at top (t), bridge (b), and hollow (h) adsorption sites (see
Figure~\ref{fig:100Q}).  As with the (111) surfaces, we consider
adsorbates at this coverage to be isolated.  

At $\Theta$=0.5 on (100) surfaces, we
modeled chemisorption at two different top site configurations, tt1 and
tt2, two different hollow site configurations, hh1 and hh2, and three
different bridge site configurations, bb1, bb2, and bb3.  These
overlayer patterns are shown in Figures~\ref{fig:100Half1},
\ref{fig:100Half2}, and \ref{fig:100Bb3}.  The top sites vary in
that tt1 has CO molecules at nearest neighbor distances, while tt2
does not.  Likewise, the hh1 overlayer has adsorbates at nearest
neighbor separations while the hh2 overlayer does not.  
In the bridge site structures, both bb1 and bb2
have CO molecules at nearest neighbor separations, but differ in that
bb2 has bonding competition while bb1 does not.  The bb3 pattern is
experimentally observed on both Pt and
Rh~(100).~\cite{Martin95p69,Baraldi96p1}  We also modeled tttt and
hhhh adsorption at $\Theta$=1.  All modeling of CO on the
(100) surfaces was done in a $c(4\times2)$ cell.

\section{\label{sec:results} Results}

We begin with low coverage ($\Theta$=0.25) results.  Detailed
comparison of these calculated energies and experimental results have
been reported previously.~\cite{Mason04p161401R} In the low-coverage
limit, our calculations reproduce the site preferences observed
experimentally for Rh, Pd and Pt~(111) surfaces (top for Rh and Pt, fcc for
Pd).~\cite{Wei97p49} Experimental data show that the lowest-coverage
ordered CO structure on all three of these (111) surfaces is
$(\sqrt{3}\times\sqrt{3})R30^\circ$ at $\Theta$=0.33.  The formation
of this overlayer is driven by
nearest neighbor CO-CO repulsive interactions, since this overlayer 
gives the highest-coverage single-site occupation 
without nearest neighbor interactions. The
site occupied in the $(\sqrt{3}\times\sqrt{3})R30^\circ$ structure is
determined by the site preference of isolated CO adsorption.

The (100) surfaces of Pt, Rh, and Pd
share common low-coverage chemisorption behavior.  The lowest coverage
at which ordered CO structures are observed on the (100) surfaces is
$\Theta$=0.5.  For (100) surfaces, we find that difference in $E_{\rm
chem}$ between the $\Theta$=0.25 CO overlayer and the highest $E_{\rm
chem}$ $\Theta$=0.5 CO overlayer is negligible, in agreement with
experiments that find a $c(4\times2)$ overlayer at $\Theta$=0.5 to
be the lowest coverage ordered structure.~\cite{Wei97p49}  On Pt~(100), 
domains of top-site CO and of bridge-site CO are seen.~\cite{Wei97p49}
On Rh~(100), CO occupies top sites at $\Theta$=0.5, and
on Pd~(100) CO binds at bridge sites.~\cite{Wei97p49}  For
all three metals, our results for site preference at $\Theta$=0.25 are
consistent with these experimental observations.

While Pd, Rh and Pt~(111) surfaces form the same ordered structure at
low coverage, they evolve differently with coverage.
Comparison between DFT and experiment at higher $\Theta$ is imprecise
due to our neglect of temperature, pressure, and entropy effects.
However, our DFT results for $\Theta$=0.5 and 0~K are consistent with
experimental results.  Our calculations show that chemisorption of CO
on Pt~(111) at $\Theta$=0.5 in the $c(4\times2)$ tb mixed occupation
structure is favored by $\approx0.09$~eV/molecule over the next highest $E_{\rm
chem}$ overlayer, in agreement with the experimental observations of
the $c(4\times2)$ tb structure at
$\Theta$=0.5.~\cite{Pederson99p403,Ogletree86p351,Bondino00pL467}  For
Pd~(111) we find that the bb3 $c(4\times2)$ structure is favored by
$\approx0.04$~eV over the $c(4\times2)$ hf structure.  All other
overlayer patterns show $E_{\rm chem}$ values more than 0.1~eV lower
than the hf and bb3 structures.  This result is in agreement with
coexistence of $c(4\times2)$ bb3 and hf overlayers at
$\Theta$=0.5 observed on Pd~(111) in low-temperature STM
experiments.~\cite{Rose02p48}  

The case of Rh~(111) is especially interesting, with the experimental
LEED pattern showing disorder at
$\Theta$=0.5.~\cite{Beutler98p117,Beutler97p381} Our results show that
the bb3, tb, and hf overlayers on Rh~(111) at $\Theta$=0.5 have the
same $E_{\rm chem}$ to within 0.03~eV.  The experimental result may be
due to full disorder, but our computations lead us to suggest that
domains of several nearly isoenergetic overlayer patterns coexist,
explaining the complicated LEED pattern.

For Pt~(100) and
Rh~(100) surfaces, we find a small energy
difference between the top and bridge site, consistent with the
occupation of both of these sites for $\Theta$=0.5 and
greater.~\cite{Martin95p69,Baraldi96p1}  For Pd~(100), the large site
preference energy is consistent with experimental observation that
only bridge sites are occupied for all values of
$\Theta$.~\cite{Bradshaw78p513}

\begin{table}
\caption{Results for CO $E_{\rm chem}$ at different $\Theta$ on the
(111) surfaces of Pt, Rh, and Pd, in eV.  Selected Al values are
also reported.  For coverages above $\Theta$=0.25, the difference $E_{\rm int}$
between the low-coverage value and high-coverage value $E_{\rm chem}$ for the
corresponding sites is reported in parenthesis.  Values for
experimentally observed structures are in italics.  *~indicates systems
for which the lateral forces on CO were not minimized (CO is
constrained to be perpendicular in all cases).}

\begin{tabular}{lrr|rr|rr|rr}
&\multicolumn{2}{c}{Pt~(111)}
&\multicolumn{2}{c}{Rh~(111)}
&\multicolumn{2}{c}{Pd~(111)}
&\multicolumn{2}{c}{Al~(111)}\\
              & $E_{\rm chem}$& $E_{\rm int}$&
$E_{\rm chem}$& $E_{\rm int}$&$E_{\rm chem}$& $E_{\rm int}$&
$E_{\rm chem}$& $E_{\rm int}$\\
\hline
$\Theta$=0.25 & & & & & & & & \\
t       &   {\em 1.56}  & & {\em 1.67}  & & 1.25 & & 0.22\\
b       &   1.43        & & 1.58        & & 1.49 & & \\
h       &   1.40        & & 1.64        & & 1.60 & & \\
f       &   1.43        & & 1.64        & & {\em 1.63} & & \\
\hline
$\Theta$=0.5 & & & & & & & & \\
tt      &   1.27 & 0.29   & 1.35 & 0.32   & 0.93 & 0.32  & 0.17 & 0.05 \\
bb1*    &   1.12 & 0.31   & 1.25 & 0.33   & 1.15 & 0.34 \\
bb2     &   1.07 & 0.36   & 1.40 & 0.18   & 1.09 & 0.40 \\
bb3*    &   1.36 & 0.07   & {\em 1.55} & 0.03 & {\em 1.44} & 0.05 \\
tb*     & {\em 1.45} & 0.04 & {\em 1.56} & 0.06 & 1.31 & 0.06 & \\
hh*     &   1.03 & 0.37   & 1.33 & 0.31   & 1.19 & 0.41 \\
hf      &   1.21 & 0.20   & {\em 1.53} & 0.11 & {\em 1.40} & 0.21 \\
\hline
$\Theta$=0.75 & & & & & & & & \\
thf*    &   1.19 & 0.27   & {\em 1.39} & 0.26 & {\em 1.15} & 0.34 \\   
\hline
$\Theta$=1 & & & & & & & & \\
tttt    &   0.74 & 0.82   & 0.53 & 1.14 & 0.39 & 0.86 \\
hhhh*   &   0.31 & 1.09   & 0.76 & 0.88 & 0.46 &1.14  \\
\hline
\end{tabular}
\label{table:111Results}
\end{table}

\begin{table}
\caption{Same as Table~\ref{table:111Results}, for CO chemisorption on
(100) surfaces.}

\begin{tabular}{lrr|rr|rr|rr}
&\multicolumn{2}{c}{Pt~(100)}
&\multicolumn{2}{c}{Rh~(100)}
&\multicolumn{2}{c}{Pd~(100)}
&\multicolumn{2}{c}{Al~(100)}\\
              & $E_{\rm chem}$& $E_{\rm int}$&
$E_{\rm chem}$& $E_{\rm int}$&$E_{\rm chem}$& $E_{\rm int}$&
$E_{\rm chem}$& $E_{\rm int}$\\
\hline
$\Theta$=0.25 & & & & & & & & \\
t       &   {\em 1.80}  & & 1.72  & & 1.35 & & 0.12 \\
b       &   {\em 1.83}  & & {\em 1.77}  & & {\em 1.64} & \\
h       &   1.25        & & 1.60        & & 1.50 & \\
\hline
$\Theta$=0.5 & & & & & & & & \\
tt1     &   1.49 & 0.31 & 1.39 & 0.33  & 1.01 & 0.34 & 0.07 & 0.05 \\
tt2     &   {\em 1.78} & 0.02 & {\em 1.72} & 0.00 & 1.34 & 0.01 \\
bb1     &   1.60 & 0.23 & 1.47 & 0.30  & 1.36 & 0.28 \\
bb2     &   1.50 & 0.33 & 1.58 & 0.19  & 1.28 & 0.36 \\
bb3     & {\em 1.77} & 0.06 & {\em 1.73} & 0.04 & {\em 1.62} & 0.02 & \\
hh1     &   0.83 & 0.42 & 1.23 & 0.37 & 1.10 & 0.40 \\
hh2     &   1.08 & 0.17 & 1.49 & 0.11 & 1.35 & 0.15 \\ 
\hline
$\Theta$=1 & & & & & & & &  \\
tttt    &   1.16 & 0.64 & 1.08 & 0.64  & 0.70 & 0.65 \\
hhhh    &   0.11 & 1.14 & 0.73 & 0.87  & 0.38 & 1.12 \\ 
\hline
\end{tabular}
\label{table:100Results}
\end{table}

\section{\label{sec:discuss} Discussion}

In this section, we discuss trends in the data presented in
Tables~\ref{table:111Results} and \ref{table:100Results}.  We
identify the presence of nearest neighbors and/or metal-atom sharing as the
chief determinants of interactions between adsorbed CO molecules.  

The effects of the substrate on adsorbate-adsorbate interactions will be
discussed in the framework of the Hammer-Morikawa-N\o rskov
model~\cite{Hammer96p2141} for the $d$-band contribution to
top-site CO chemisorption:

\begin{eqnarray}
E_{\rm chem}^{d}= 4\left[\frac { f V_{\pi}^2}{(\epsilon_{2\pi^*}-\epsilon_d)}+fS
_{\pi}V_{\pi}\right]\nonumber\\+2\left[(1-f)\frac{V_{\sigma}^{2}}{(\epsilon_{d}-\epsilon_{5\sigma})}
+(1+f)S_{\sigma}V_{\sigma}\right]
\label{eqn:HNmodel}
 \end{eqnarray}
\noindent where $f$ is the idealized filling of the metal $d$ bands,
$V_{\pi}^2$ is the metal-carbon orbital overlap matrix element at the
top site,
$\epsilon_{2\pi^*}$ and $\epsilon_{5\sigma}$ are the energies of the
renormalized 2$\pi$* and 5$\sigma$ orbitals relative to the Fermi
level, and $\epsilon_d$ is the center of the metal $d$-bands relative
to the Fermi level.  $V$ and $S$ are coupling and overlap matrix
elements respectively, and are labeled by orbital.

We also analyze gas-phase CO-CO interactions, study induced charge
densities in chemisorbed systems, and analyze the metal electronic
structure.

\subsection{\label{sec:through} 
Nearest-neighbor interactions and
chemisorption-induced perturbations}

We now analyze the $E_{\rm chem}$ results presented in Section III to
understand nearest-neighbor adsorbate-adsorbate interactions, for
systems that have no metal-atom sharing.

We can estimate the energetic cost of placing adsorbates at nearest
neighbor distances from our top-site chemisorption results.  Our
dataset includes three different exclusively top site chemisorbed
structures at $\Theta$=0.5: The tt structure on (111) (see
Figure~\ref{fig:111TH}), and the tt1 and tt2 structures on the (100)
surface (see Figures~\ref{fig:100Half1} and \ref{fig:100Half2}).
While the chemisorbed CO molecules in the tt and tt1 structures have
nearest-neighbor CO adsorbates, the adsorbates in the tt2 structure do
not.  From inspection of the data in Tables~\ref{table:111Results} and
\ref{table:100Results}, it is clear that $E_{\rm int}$ (the difference
between the low-coverage limit and higher coverage $E_{\rm chem}$) for
the tt2 structure on the (100) surface is tiny ($\le$0.02~eV) for all
metals.  $E_{\rm int}$ is much larger but still nearly independent of
metal identity for the tt and tt1 patterns (=0.32~eV$\pm$0.02~eV).
$E_{\rm int}$ for the Al surfaces is almost an order of magnitude
smaller than for the transition metal surfaces.

We can compare these $E_{\rm int}$ results 
with the gas-phase CO-CO interaction
energy $E_{\rm int}^{\rm gas}$, calculated by:

\begin{eqnarray}
E_{\rm int}^{\rm gas}(r) = E^{\rm gas}_{\rm CO-CO} - 2E_{\rm CO}   
\end{eqnarray}
\begin{figure}
\includegraphics[width=3.0in]{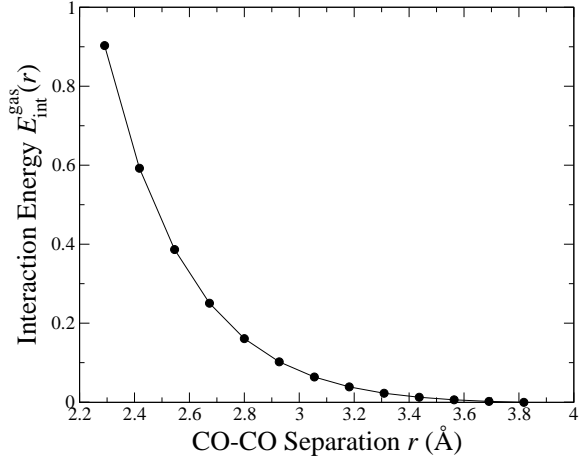}
\caption{{Calculated interaction energy $E_{\rm int}^{\rm gas}(r)$ for 2 CO molecules at different separations $r$.}}
\label{fig:COint}
\end{figure}

The interaction energy for CO at different separations is plotted in
Figure~\ref{fig:COint}.  At the nearest-neighbor separations for Al
(2.89~\AA), Pt (2.76~\AA), Pd (2.73~\AA), and Rh (2.66~\AA), the gas-phase
CO-CO interaction energies are 0.12, 0.19, 0.21, and 0.27~eV,
respectively.  The corresponding (111) tt $E_{\rm int}$ values 
from Table~\ref{table:111Results} are 
0.05, 0.30, 0.32, and 0.32~eV,
respectively.  
For Al,
$E_{\rm int}<E_{\rm int}^{\rm gas}$,
suggesting that the $s$ and $p$ metal electrons do
not increase CO-CO repulsion and perhaps lessen it somewhat.
The transition metals all show 
$E_{\rm int}>E_{\rm int}^{\rm gas}$,
meaning that through-space repulsion is augmented by the surface.
From the contrast with Al,
we deduce that
this increase in repulsion must be due to $d$ electrons 
changing
the adsorbed CO
electronic structure.  

The Hammer-Morikawa-N\o rskov model 
(Equation~\ref{eqn:HNmodel}) shows that the back-donation from the metal
$d$-bands into the CO 2$\pi^*$ 
increases with $f$.  Increased filling of the
2$\pi^*$ orbitals strengthens the repulsion between chemisorbed CO
molecules, while increased inter-adsorbate separation weakens it.
Pt and Pd have a higher $f$ 
but a greater CO-CO separation than on Rh.  
Our data indicate that these effects
balance, giving nearly the same CO-CO interaction energy
(top-site $E_{\rm int}$)
for these three transition metals.

The bridge-site systems show the same $E_{\rm int}$ trends.  For 
systems with no first nearest neighbors (bb3) $E_{\rm int}$ is very
small.  For bridge overlayers at $\Theta$=0.5 with nearest neighbors
but no metal-atom sharing (bb1), a large metal-independent $E_{\rm
  int}\approx 0.32$~eV is observed, just as for top sites.  (Actually,
the bb1 pattern on Pt~(100) is somewhat more stable; this surface is
known to reconstruct, so chemisorption may be stabilized by relieving
surface
stress.~\cite{Brown98p797,Martin95p69,Morgan68p405,Morgan69p3309}  The changes in $E_{\rm chem}$ we do observe on the unreconstructed Pt(100) surface are in excellent agreement with other recent theoretical work.~\cite{Yamagishi05p8899})

Nearest-neighbor hollow sites always involve metal atom sharing, so
they are considered in the next section.

\subsection{\label{sec:bondcomp} Metal-atom sharing:  bonding competition
and electron delocalization}

Bonding competition arises when the surface metal atoms involved in
chemisorption participate in more than one carbon-metal bond.  To find the
susceptibility of $E_{\rm chem}$ to external perturbation (such as the
sharing of metal atoms) within the
Hammer-Morikawa-N{\o}rskov model, we differentiate the expression for
$E_{\rm chem}^d$ (Equation~\ref{eqn:HNmodel}) with respect to external
parameter $\lambda$, representing the number of
chemisorption bonds per metal atom:

\begin{eqnarray}
\frac{  d E_{\rm chem}^d}{d \lambda} =  \left[\frac{4f V_{\pi}^2} {
(\epsilon_{2\pi^*}-\epsilon_{d})^{2}}
- 
\frac{2(1-f)V_{\sigma}^{2}} {(\epsilon_{d}-\epsilon_{5\sigma})^{2} } \right]\frac{d \epsilon_{d}}{d \lambda}
\label{eqn:deriv}
\end{eqnarray}

The opposite signs of the two terms in Equation~\ref{eqn:deriv} mean
that any change in the position of $\epsilon_{d}$ has two competing
effects on $E_{\rm chem}$.  This competition vanishes as $f\rightarrow
1$, so $E_{\rm chem}$ on Pd or Pt ($f=$0.9) should be more sensitive
to bonding competition than Rh ($f=$0.8).

The change of $E_{\rm chem}$ with bonding competition $\lambda$ also
depends on $d\epsilon_d/d\lambda$, the shift in the $d$-band energy as
chemisorption bonds are formed.  
The $\epsilon_d$ position is affected by perturbation to the metal
surface such as strain~\cite{Mavrikakis98p2819,Wu00p1177},
alloying~\cite{Hammer96p2141,Greeley04p810} and creation of chemical
bonds with adsorbates.~\cite{Hammer01p205423,Ruban97p421}  
This effect is explored in
Table~\ref{table:Bb100DBC}, showing that
CO chemisorption causes a much smaller shift in $\epsilon_d$ for Rh than
for either Pd or Pt.  

\begin{table}
\caption{Values for the $d$-band center, $\epsilon_{d}$, 
for the (111) and (100) surfaces
of Pt,
Rh, and Pd when no CO is adsorbed ($\Theta$=0), and when CO is
adsorbed at bridge site at $\Theta$=0.25.  
The changes in $\epsilon_{d}$ due to 
chemisorption ($\Delta$) are also listed.  All reported values are in eV.}

\begin{tabular}{lrccc}
&&\multicolumn{1}{c}{$\Theta$=0}
&\multicolumn{1}{c}{$\Theta$=0.25}
&\multicolumn{1}{c}{$\Delta$}\\

\hline
Pt&(111)  &  -2.93  &  -3.59  & -0.66\\
Pt&(100)  &  -2.79  &  -3.50  & -0.71\\
Pd&(111)  &  -2.29  &  -3.04  & -0.75\\
Pd&(100)  &  -2.17  &  -2.94  & -0.77\\
Rh&(111)  &  -2.42  &  -2.85  & -0.43\\
Rh&(100)  &  -2.29  &  -2.73  & -0.44\\

\end{tabular}
\label{table:Bb100DBC}
\end{table}

The preceding analysis of the Hammer-Morikawa-N{\o}rskov model
suggests that bonding competition should increase $E_{\rm int}$ for
all chemisorption involving metal sharing, but much less on Rh than on
Pt or Pd, due to lower $f$ and lower $d\epsilon_d/d\lambda$.

\begin{figure}
     \centering
     \subfigure[]{
          \label{fig:ChdB1Top}
          \includegraphics[width=3in]{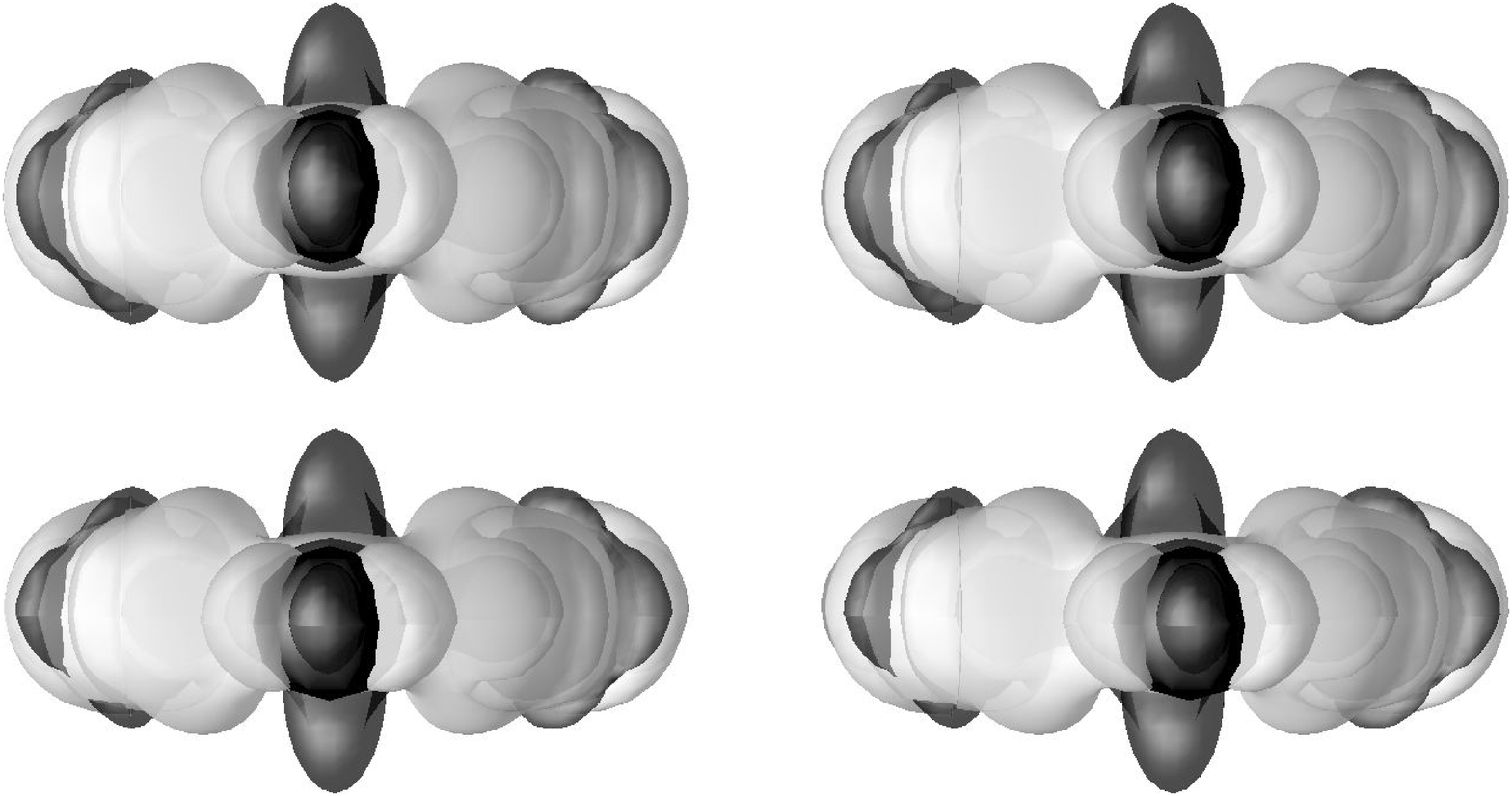}}
     \hspace{.3in}
     \subfigure[]{
          \label{fig:ChdB1Side}
          \includegraphics[width=3in]{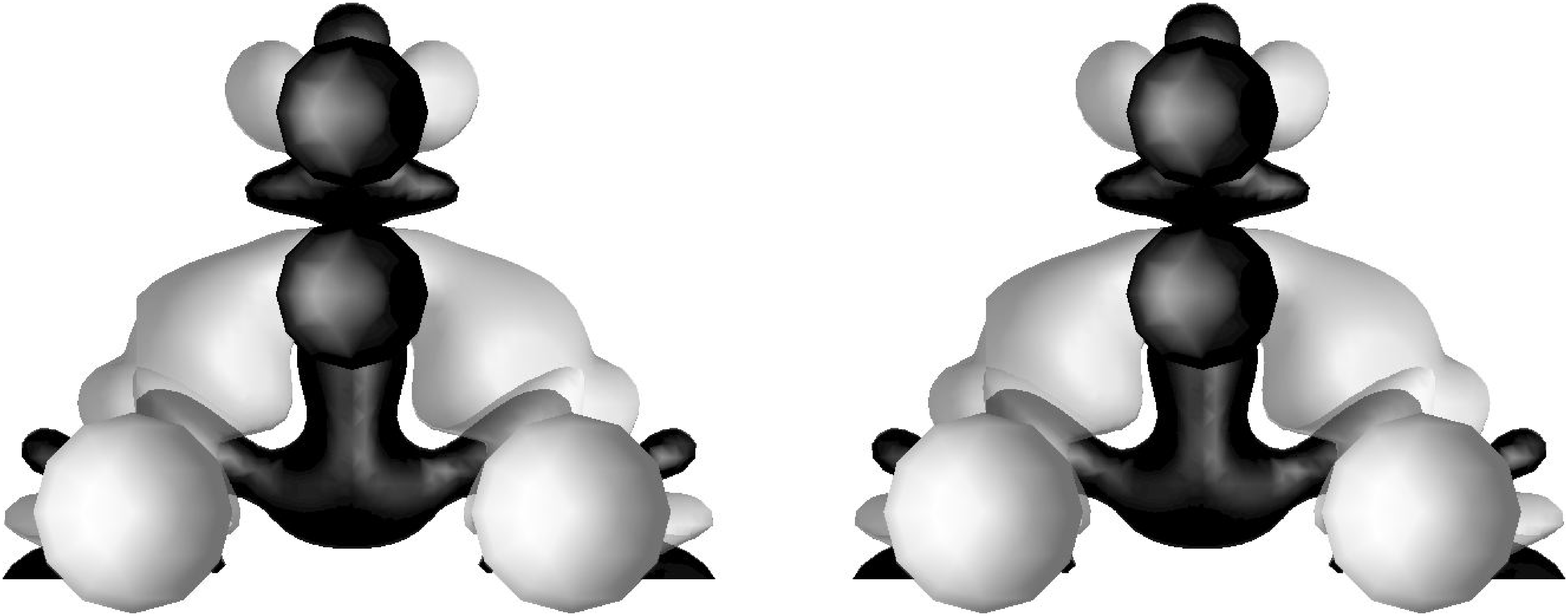}}\\
     \caption{Charge density changes induced by CO adsorption in pattern bb1
     site on Pt~(100).  Lightly shaded iso-surfaces show areas of charge
     gain, and darker iso-surfaces show areas of charge loss.  (a) 
     Top view. (b) Side view.}
     \label{fig:ChdB1}
\end{figure}

\begin{figure}
     \centering
     \subfigure[]{
          \label{fig:ChdB2Top}
          \includegraphics[width=3in]{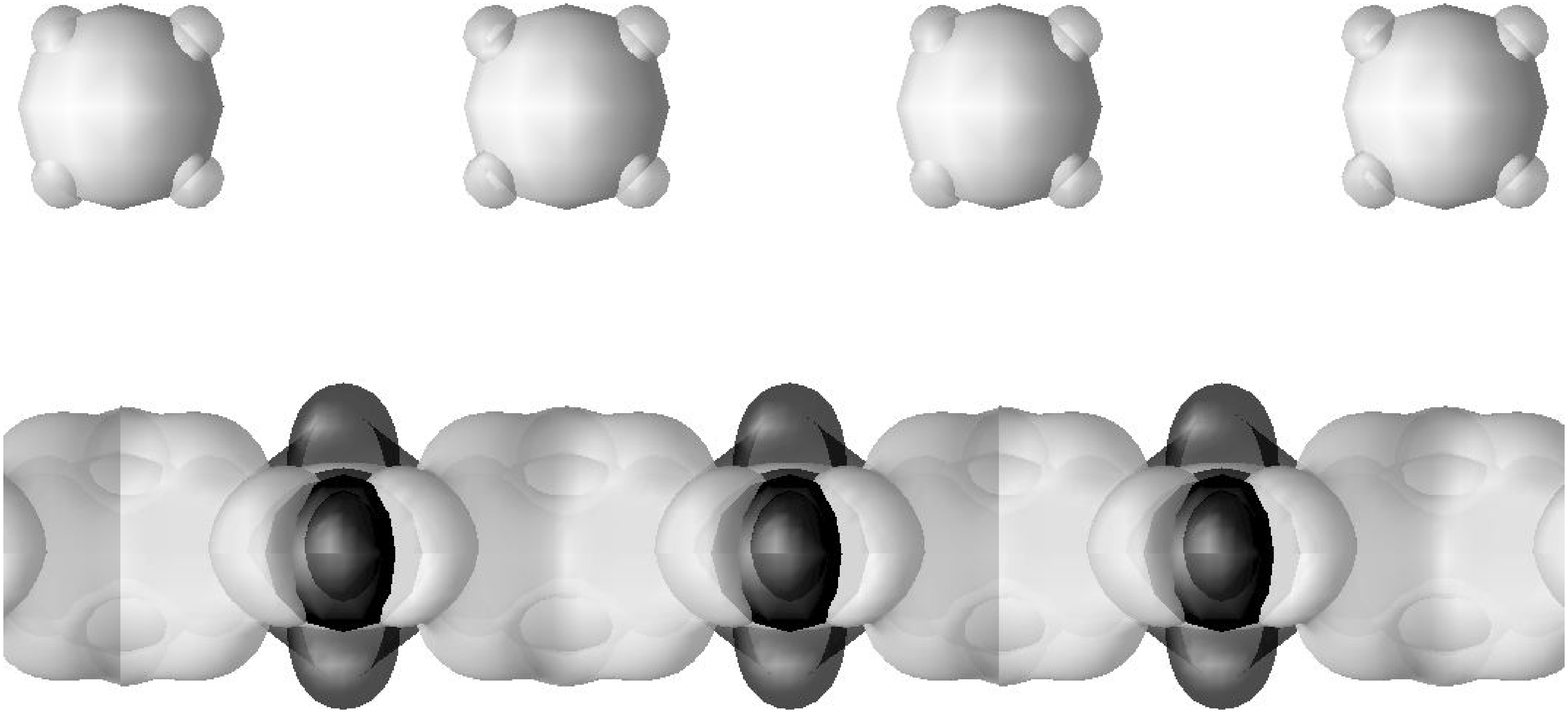}}
     \hspace{.3in}
     \subfigure[]{
          \label{fig:ChdB2Side}
          \includegraphics[width=3in]{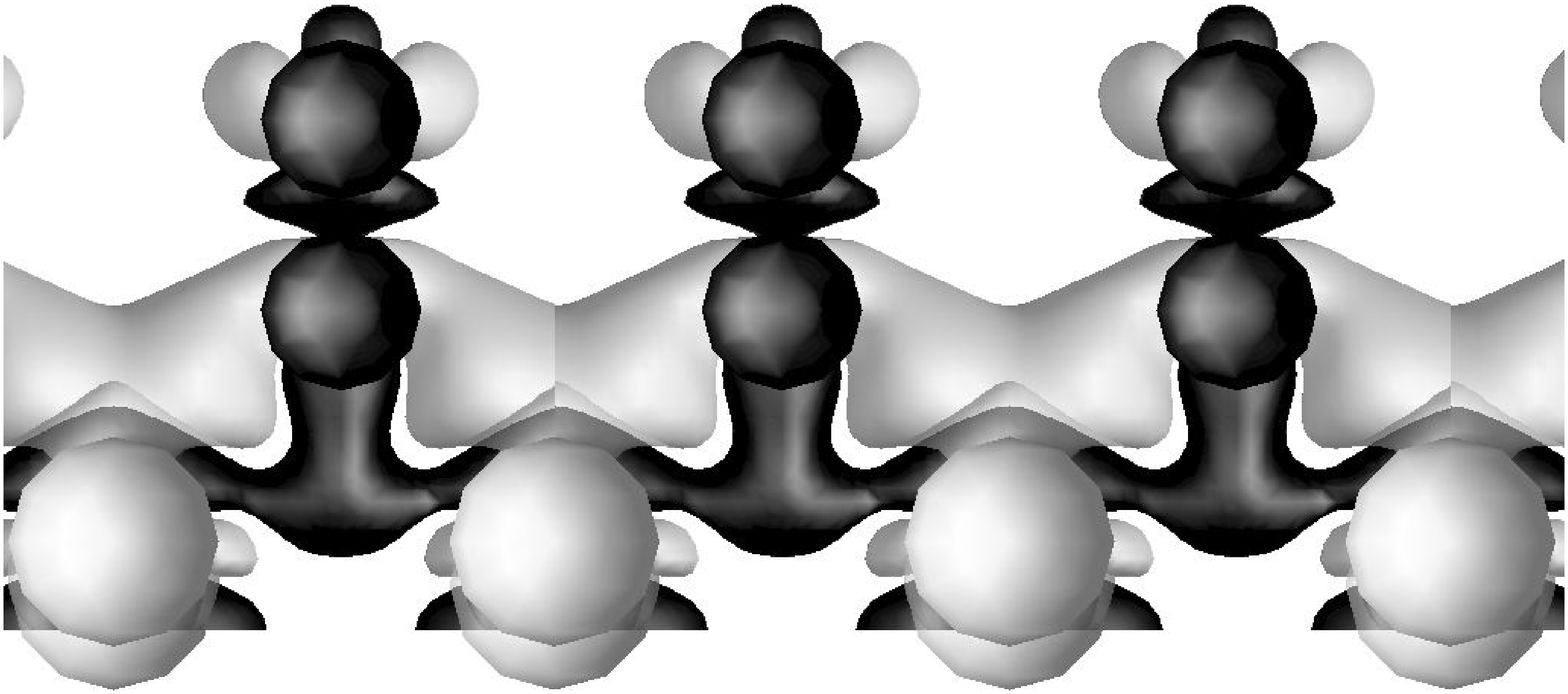}}\\
     \caption{Same as Figure~\ref{fig:ChdB1}, for the bb2 overlayer.  Note the significant charge sharing between neighboring CO molecules, mediated by the shared metal atom.}
       \label{fig:ChdB2}
\end{figure}

In addition to the destabilizing effect of bonding competition,
metal-atom sharing also enables electron delocalization.  As
illustrated in Figures~\ref{fig:ChdB1} and \ref{fig:ChdB2}, the CO
2$\pi^*$ orbitals of neighboring molecules and the metal $d$ orbitals
of the atom they share combine to form extended states in the bb2
system (Figure~\ref{fig:ChdB2}).  This delocalization is not present
on the bb1 overlayer(Figure~\ref{fig:ChdB1}).

For the bridge-bonded systems with metal-atom sharing (bb2), the
effect is dramatic.  On Rh~(111) and (100), delocalization stabilizes
chemisorption significantly, and bonding competition gives a only
small reduction, leading to stronger chemisorption and $E_{\rm int}$
values that are 0.1--0.15~eV lower for bb2 than for bb1.  On Pt and
Pd, much stronger bonding competition destabilizes chemisorption,
raising $E_{\rm int}$ for bb2 about 0.15--0.2~eV above that of Rh, and
0.05--0.1~eV above the bb1 values on Pt and Pd despite electron
delocalization.

Hollow site chemisorption exhibits the same trends, with bonding
competition making bonding less effective, and metal-mediated
delocalization stabilizing metal-sharing chemisorption.  However,
because a CO at a hollow site distributes its bonding among three or
four surface atoms, both the bonding competition and delocalization
effects from metal sharing are somewhat less than for bridge sites.
In all cases (hh,hh1,hh2) $E_{\rm int}$ for Rh is more favorable by
$\approx$0.05~eV than on Pt and Pd.  Since $E_{\rm int}$ at hh and hh1
is greater than at tt and tt1, metal sharing appears to always be net
repulsive for hollow sites.

\subsection{\label{sec:highcov} Preferred Overlayer Structures}

In this section we discuss how 
adsorbate-adsorbate interactions, along with the low-coverage CO site
preference on each surface, influence $E_{\rm chem}$ for
various $\Theta$=0.5 overlayers.

On the (111) surfaces of Pt, Pd, and Rh, we modeled chemisorption at
$\Theta$=0.5 for seven different overlayer patterns.  In the case of
Pt~(111), we find tb (experimentally observed on
Pt~(111)~\cite{Pederson99p403,Ogletree86p351,Bondino00pL467}) has the
highest $E_{\rm chem}$, by at least 0.09~eV over the other overlayer
patterns.  At low coverage, t is preferred by 0.13~eV over b and
f, and by 0.16~eV over h.  On the (111) surface, exclusive occupation of
top sites (at $\Theta\ge$0.5) must involve nearest-neighbor adsorbates,
whereas the mixed occupation tb avoids nearest neighbors, reducing
through-space repulsion.  Therefore, the tb structure can be viewed as
a compromise between the
site preference for t, and reduced adsorbate-adsorbate repulsion
made possible by partial b occupation.  

On Pd~(111), we find $E_{\rm
chem}$ for bb3 and hf (both of which are experimentally
observed~\cite{Rose02p48}) to be within 0.04~eV of each other, and at
least 0.09~eV stronger than the other five patterns.  At low
coverage, we find f to be preferred by 0.14~eV over b.  Here we can
also interpret the hf and bb3 overlayer patterns to be compromises
between inherent site preference and adsorbate-adsorbate repulsion
minimization.  In the bb3 pattern, the occupation of less-favored
bridge sites is compensated by reduced adsorbate-adsorbate
through-space repulsion and absence of metal sharing.  Conversely,
occupying the preferred hollow sites in the hf arrangements
comes at the
expense of metal sharing, which we have discussed as being unfavorable
for Pd.  

On Rh~(111) we find bb3, tb, and hf overlayer
patterns to have the same $E_{\rm chem}$ within 0.03~eV.  At low
coverage, we find that Rh has a relatively weak inherent site preference,
preferring t over b by 0.09~eV, and t over h and f by only 0.03~eV.
Since both site preference and repulsion due to metal sharing are
weak on Rh, we can explain the stronger
$E_{\rm chem}$ of bb3, tb, and hf as being due to 
not having any nearest-neighbor CO distances.
The other four overlayers studied all have some nearest neighbors.

On the (100) surfaces of Pt, Pd, and Rh, mixed site occupation is not
observed at $\Theta$=0.5.  The square lattice geometry of the (100)
surface permits single-site occupancy at $\Theta$=0.5
without nearest neighbors.
As a result (contrary to the (111) surface), 
a tb arrangement on the (100) surface would introduce
adsorbates at shorter separations than in the tt2 arrangement.
Therefore, on (100) surfaces we would expect overlayer patterns that
minimize through-space repulsion and free of bonding competition most
likely to be observed.  Both tt2 and bb3 minimize adsorbate-adsorbate
through-space repulsions and do not involve bonding competition, and
this explains why we find $E_{\rm int}$ to be no greater than 0.06~eV
for these patterns on all three metals.

\section{\label{sec:conclude} Conclusion}    
We have presented for the first time results for CO chemisorption at
different coverages using DFT-GGA plus our first-principles
extrapolation procedure for accurate chemisorption energies.  Our
values for $E_{\rm chem}$ at higher coverage with CO occupying
different sites and in different patterns show the experimentally
observed structure to have the most favorable $E_{\rm chem}$ within
the selection of adsorbate arrangements modeled.  We identify and
dicuss adsorbate through-space repulsion, bonding competition, and
substrate-mediated electron delocalization as key factors determining
preferred chemisorption patterns for different metal surfaces and
adsorbate coverages.  We rationalize how these effects, along with the
inherent site preference energy for CO on each metal, balance to cause
different trends in chemisorption behavior as a function of CO
coverage $\Theta$ on different transition metal surfaces.

\section{\label{sec:ack} Acknowledgments}    
This work was supported by the Air Force Office of Scientific
Research, Air Force Materiel Command, USAF, under grant number
FA9550--04--1--0077, and the NSF MRSEC Program, under Grant
DMR05--20020.  Computational support was provided by the Defense
University Research Instrumentation Program, the High Performance
Computing Modernization Office of the Department of Defense, and by
the National Science Foundation CRIF Program, Grant CHE--0131132.  SEM
thanks Alexie M. Kolpak and Valentino R. Cooper for testing and
discussions of dipole corrections to chemisorption system energies.

\bibliography{Submit1123}
\end{document}